\documentclass[iop,twocolappendix,numberedappendix]{emulateapj}
\usepackage{epsfig}
\usepackage{graphics,graphicx}
\usepackage{tikz}
\usepackage{subfigure}
\usepackage{multirow}
\usepackage{amssymb}
\usepackage{natbib}
\usepackage{rotating}
\usepackage{setspace}
\usepackage[bookmarks=false, colorlinks=true, citecolor=blue, backref=true]{hyperref}
\usepackage{apjfonts}
\newcommand{\kms}{${\rm km\;s^{-1}}$}

\usepackage{graphicx}
\usepackage{color}
\usepackage{wrapfig}

\shorttitle{NOEMA CO Observations of Arp~142 \& Arp~238}
\shortauthors{C.K. Xu et al.}

\newcommand{\lsim}{\; \lower2truept\hbox{${< \atop\hbox{\raise4truept\hbox{$\sim$}}}$}\;}
\newcommand{\gsim}{\; \lower2truept\hbox{${> \atop\hbox{\raise4truept\hbox{$\sim$}}}$}\;}

\begin{document}

\slugcomment{{\bf Accepted Version}; \today}

\title{NOEMA Observations of CO Emission in Arp~142 and Arp~238}
\author{
C.~K.~Xu\altaffilmark{1,2}, 
U.~Lisenfeld\altaffilmark{3,4},
Y.~Gao\altaffilmark{5,6},
F.~Renaud\altaffilmark{7}
}
\email{Email: congxu@nao.cas.cn}
\altaffiltext{1}{Chinese Academy of Sciences South America Center for Astronomy, National Astronomical Observatories, CAS, Beijing 100101, China}
\altaffiltext{2}{National Astronomical Observatories, Chinese Academy of Sciences (NAOC), 20A Datun Road, Chaoyang District, Beijing 100101, China}
\altaffiltext{3}{Dept. F\'isica Te\'orica y del Cosmos, Campus de Fuentenueva, Edificio Mecenas, Universidad de Granada, E-18071 Granada}
\altaffiltext{4}{Instituto Carlos I de F\'isica T\'orica y Computacional, Facultad de Ciencias, E-18071 Granada, Spain}
\altaffiltext{5}{Department of Astronomy, Xiamen University, 422 Siming South Road, Xiamen 361005, China}
\altaffiltext{6}{Purple Mountain Observatory \& Key Laboratory for Radio Astronomy, Chinese Academy of Sciences, 10 Yuanhua Road, Nanjing 210023, China}
\altaffiltext{7}{Department of Astronomy and Theoretical Physics, Lund Observatory, Box 43, SE-221 00 Lund, Sweden}

\received{March~14, 2021}
\accepted{June~27, 2021}

\begin{abstract}
Previous studies have shown significant differences in the enhancement 
of the star-formation rate (SFR) and the star-formation efficiency 
($\rm SFE = SFR/M_{mol}$) between spiral-spiral and spiral-elliptical mergers. 
In order to shed light on the physical mechanism of these differences, 
we present NOEMA observations of the molecular gas distribution and kinematics
(linear resolutions of $\rm \sim 2\; kpc$)
in two representative close major-merger star-forming pairs: 
the spiral-elliptical pair Arp~142 and the
  spiral-spiral pair Arp~238.  The CO in Arp~142 is widely distributed
  over a highly distorted disk without any nuclear concentration, and
  an off-centric ring-like structure is discovered in channel maps.
  The SFE varies significantly within Arp~142, with a starburst region
  (Region~1) near the eastern tip of the distorted disk showing an SFE
  $\sim 0.3$ dex above the mean of the control sample of isolated galaxies, 
  and the SFE of the main disk (Region~4) 0.43 dex
  lower than the mean of the control sample.  In contrast, the CO
  emission in Arp~238 is detected only in two compact sources at the
  galactic centers.  Compared to the control sample, Arp~238-E shows
  an SFE enhancement of more than 1 dex whereas Arp~238-W has an
  enhancement of $\sim 0.7$ dex. We suggest that the extended CO
  distribution and the large SFE variation in Arp~142 are due to an
  expanding large-scale ring triggered by a recent high-speed head-on
  collision between the spiral galaxy and the elliptical galaxy, and
  the compact CO sources with high SFEs in Arp~238 are associated with
  nuclear starbursts induced by
  gravitational tidal torques in a low-speed coplanar interaction.
\end{abstract}


\section{Introduction}
It is well documented that mergers can trigger enhanced star-formation
in galaxies \citep{Kennicutt1987, Sanders1996}. The most extreme
starbursts, such as the ultraluminous infrared galaxies (ULIRGs:
$\rm L_{IR} \geq 10^{12} L_\sun$), are usually found in the final stage of
mergers \citep{Sanders1996}. Strong star-formation enhancements are
also detected in earlier merger stages, particularly in major-mergers
(mass ratio less than 3) during close encounters (\citealp{Xu1991};
  \citealp{Nikolic2004}; \citealp{Ellison2010}; \citealp{Scudder2012}). 
On the other hand, only a small fraction of interacting
galaxies show significant star-formation enhancement
(\citealp{Horellou1999}; \citealp{Bergvall2003}; \citealp{Knapen2009}). 
Spitzer observations of a sample of
K-band selected close major-merger pairs \citep{Xu2010}, which
preferentially select early mergers during or near the first and
second pericentric passages, found that only $\sim 25\%$ of
star-forming galaxies in the sample show strong enhancement in
specific star-formation rate 
($\rm sSFR=SFR/M_{star}$, where SFR is 
the star-formation rate in $\rm M_\sun\; yr^{-1}$ 
and $\rm M_{star}$ the stellar mass in $\rm M_\sun$).  
Furthermore, 
the far-infrared (FIR) observations by Spitzer and Herschel show that
only star-forming galaxies in spiral-spiral (hereafter S+S) 
pairs have significantly enhanced
sSFR, but not those in spiral-elliptical (hereafter S+E) 
pairs (\citealp{Xu2010}; \citealp{Cao2016}).
The low fraction
of paired galaxies with enhanced sSFR is often interpreted
as due to the fact that strong starbursts triggered by interactions are
"on" only for short periods ($\sim 100$~Myr), while most time a
merging galaxy is in the "off" phase of the 
starburst \citep{DiMatteo2008}. However,
this interpretation cannot explain the non-enhancement of sSFR 
in star-forming galaxies in 
S+E pairs, which represent 34\% of star-forming galaxies in a
complete sample of K-band selected close major-merger pairs (KPAIR,
\citealp{Domingue2009}). 
It was
suggested (\citealp{Park2009}; \citealp{Hwang2011}) that the lack of
star-formation enhancement in S+E pairs could be due to stripping
of cold gas of the spiral component by ram-pressure of the hot-gas halo
surrounding the elliptical component. But this hypothesis is rejected by the 
results of \citet{Zuo2018} and \citet{Lisenfeld2019}. \citet{Lisenfeld2019}
carried out IRAM CO observations for 78 spiral galaxies selected 
from the H-KPAIR sample of 88 close major-merger pairs that have 
Herschel FIR observations \citep{Cao2016}. 
Combining with the GBT HI observations
of \citet{Zuo2018} for pairs selected from the same H-KPAIR sample, 
\citet{Lisenfeld2019} found no significant difference between the
total gas abundances of star-forming galaxies 
in S+E and in S+S pairs. 
Indeed, their results show that the reason for spiral galaxies in S+E
pairs to have a significantly lower sSFR than their counterparts in S+S
pairs (\citealp{Xu2010}; \citealp{Cao2016}) is because they have
a significantly lower molecular-to-total-gas ratios ($\rm
M_{H_2}/(M_{H2}+M_{HI})$) and a lower star-formation efficiency 
($\rm SFE=SFR/M_{H_2}$).

In this article, we present NOEMA CO imaging observations of two
  representative pairs: Arp~142 (S+E) and Arp~238 (S+S).  In the
  sample of \citet{Lisenfeld2019}, the spiral component of Arp~142
  (NGC~2936) has the highest SFR among galaxies in S+E pairs.
  Arp~238-E (UGC8335-E), a luminous infrared galaxy (LIRG: $\rm L_{IR}
  \geq 10^{11} L_\sun$), has the second highest SFR among galaxies in
  S+S pairs (see Table~\ref{tab:results}). Interestingly, the SFE in
  Arp~142 is $\sim 30$ times lower than that in Arp~238-E
  \citep{Lisenfeld2019}.  With the NOEMA observations, we aim to probe
  the cause of the strong difference between SFEs of the two pairs,
  which may also shed light on the physical mechanism for the SFE
  difference between S+E and S+S pairs in general.

\begin{table*}
\caption{Summary of the CO(1-0) observations and properties of the final data cube.}
\label{tab:overview_obs_data}
\centering
\begin{tabular}{lll}
\noalign{\smallskip} \hline \noalign{\medskip}
 & {\bf Arp~142} & {\bf Arp~238} \\
 \noalign{\smallskip} \hline \noalign{\medskip}
{\bf Observation details: }& & \\
Dates  & April 12-15, 2020 & May 8-24, 2020 \\
Total observing time & 4.4 hours & 3.7 hours \\
Center position of mosaic  &RA: 09:37:43.90 & RA: 13:15:32.77 \\
                         &DEC: 02:45:26.0 & DEC:  62:07:37.6\\
Offsets of mosaic pointings (RA, DEC)\ \ & (7.5\arcsec, 8.0\arcsec)  & (10.0\arcsec,-5.0\arcsec) \\
      & (-7.5\arcsec,-8.0\arcsec) & (-10.0\arcsec, 5.0\arcsec)\\
Observed central frequency  & 112.644262 GHz& 111.82685  GHz\\
Flux calibrator &LKHA101 & MWC349  \\
Bandpass calibrator & 3C84 ; J0854+2006 & 3C273; J1751+0939  \\
Phase calibrator  & J0930+0034; J0909+0121 & J1302+5748 ;  J1302+6902 \\
{\bf Properties of final data cubes:} $ $ \\
Beam size of cleaned image  & 4.28\arcsec$\times$3.44\arcsec& 3.36\arcsec$\times$3.02\arcsec\\
Position angle of beam & -4\arcdeg & 88\arcdeg \\
Mean noise of cleaned image &  1.14 mJy/beam &  0.73  mJy/beam\\
(frequency resolution of 12 MHz) & & \\
\noalign{\smallskip} 
\hline \noalign{\medskip}
\end{tabular}
\end{table*}

\begin{figure*}[!htb]
\centering
\includegraphics[width=0.8\textwidth]{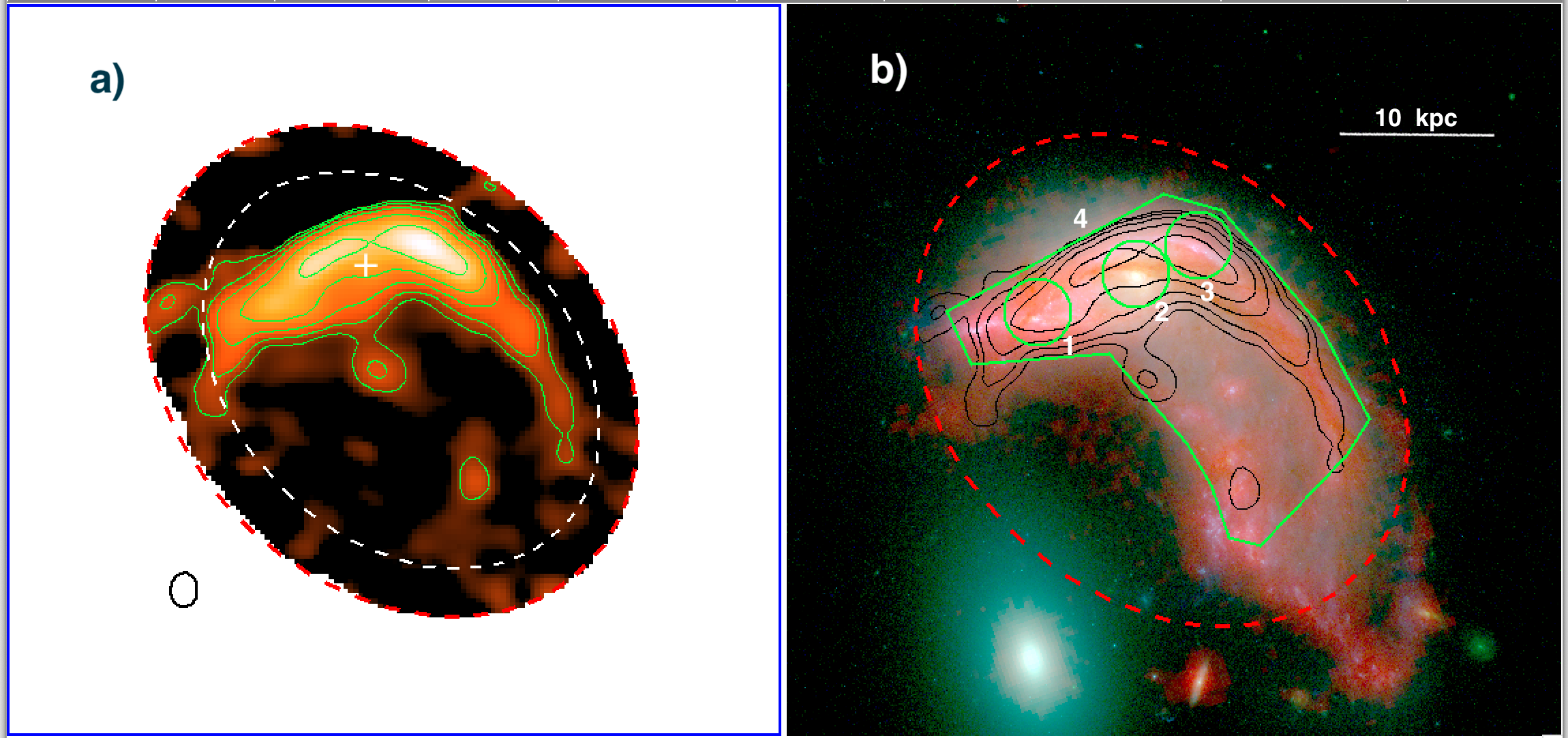}
\includegraphics[width=0.8\textwidth]{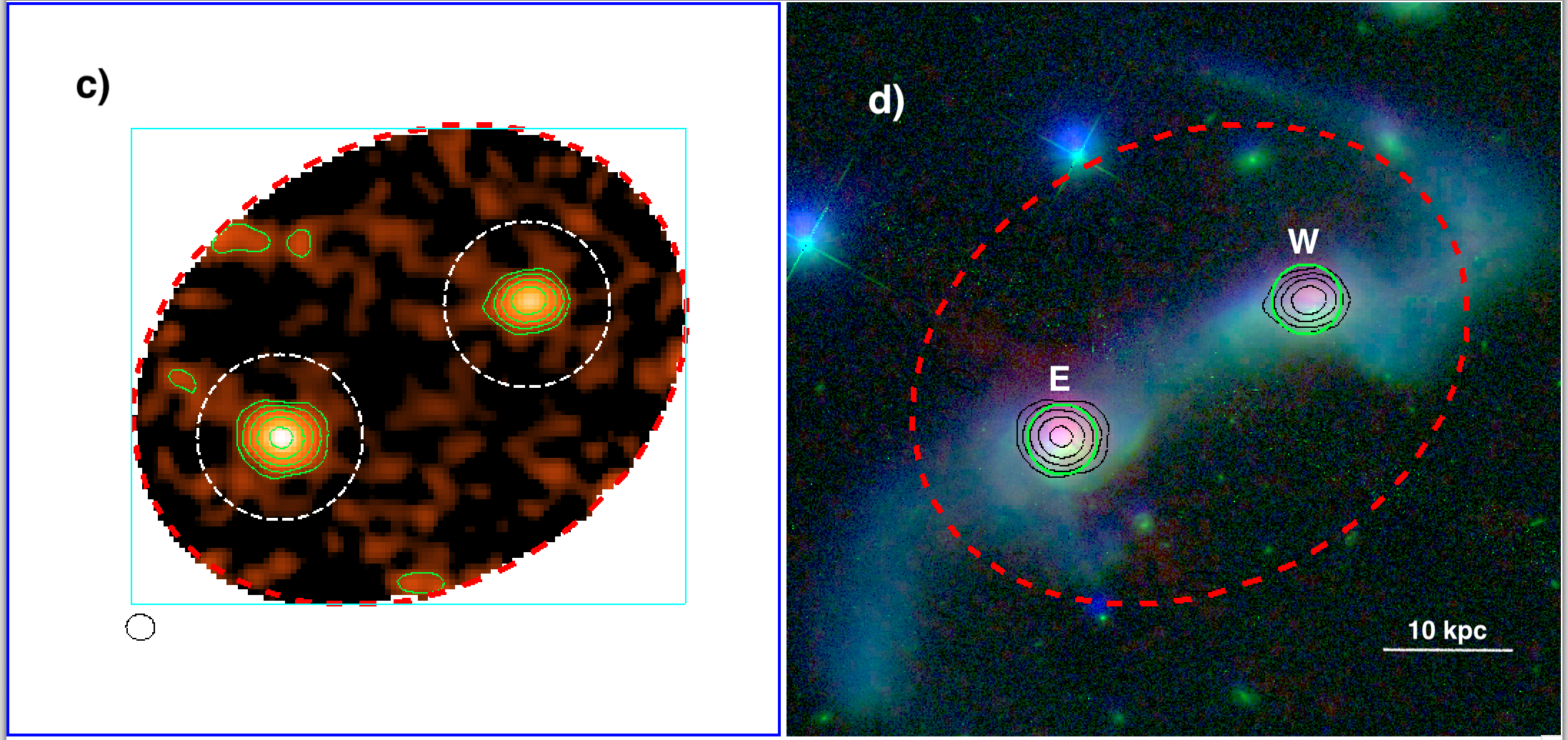}
\caption{
{\it Panel~a)}: Integrated CO(1-0) map of Arp~142. 
The contour levels are at 1, 2, 4, 8, 16
$\rm Jy\; km\; s^{-1}\; beam^{-1}$. The noise of the map is non-uniform, varying 
between 0.3 $\rm Jy\; km\; s^{-1}\; beam^{-1}$ (inner region)
and 0.5 $\rm Jy\; km\; s^{-1}\; beam^{-1}$ (outer region).
The small black ellipse in the lower-left corner of the panel represents 
the beam, which has the FWHM of 
4.28\arcsec$\times$3.44\arcsec, corresponding to a physical scale of $\rm 2.1\; kpc\times 1.7\; kpc$.
The white dashed ellipse (size=47.1\arcsec$\times$42.6\arcsec)
outlines the aperture for the measurement
of the total CO flux.  The white plus sign marks the nucleus.
{\it Panel~b)}: Optical-IR 
three-color plot (blue: HST-F475W, 
green: HST-F814W, red: IRAC-8.0$\mu m$) of Arp~142,
overlaid by the same CO contours (black) as in Panel~a).  
The four regions (with green boundaries) are: Region~1 - off-nucleus 
starburst region \citep{Mora2019},
Region~2 - nuclear region, Region~3 - CO peak region, Region~4 - main disk.  
The red dashed ellipse outlines the
field of view of NOEMA observation.
{\it Panel c)}: Integrated CO(1-0) map of Arp~238.
The  small black ellipse in the lower-left corner of the
panel represents the beam (3.36\arcsec$\times$3.02\arcsec, corresponding to a physical scale of $\rm 2.1\; kpc\times 1.9\; kpc$). The contour levels are at 
1, 2, 4, 8, 16 $\rm Jy\; km\; s^{-1}\; beam^{-1}$. The noise of the map is  
in the range of 0.2 $\rm Jy\; km\; s^{-1}\; beam^{-1}$ (inner region)
and 0.5 $\rm Jy\; km\; s^{-1}\; beam^{-1}$ (outer region). The
white dashed circles (with $D=20$\arcsec) outline the apertures for the
measurements of the total CO fluxes of the two galaxies. 
{\it Panel d)}: Optical-IR  three-color plot (blue: HST-F435W, 
green: HST-F814W, red: IRAC-8.0$\mu m$) of Arp~238, 
overlaid by the same CO contours (black) as in Panel~c).
The two green circles mark the two central regions around the two nuclei.
The red dashed ellipse outlines the field of view of NOEMA observations.
}
\label{fig:CO-IRAC}
\end{figure*}

\section{Observations}

CO(1-0) was observed with the Northern Extended Millimeter Array
(NOEMA) of the Institute of Radioastronomy in the Millimeter
(IRAM)\footnote{IRAM is supported by INSU/CNRS (France), MPG
  (Germany) and IGN (Spain).}  in C and D array with 10 antennas
(project W19BL).  The
observations were carried out under good weather conditions.  For each
object we made a small mosaic consisting of two overlapping regions.
The receiver covers two side-bands, each with a width of 7.744
GHz. The autocorrelator PolyFix was used which has a channel width of
2 MHz (corresponding to 5.3 \kms\ at the frequency of our
observations). The line frequency was tuned in the Upper Side Band
(USB). Some basic parameters of the observations are listed in
Table~\ref{tab:overview_obs_data}.

We reduced the data following standard procedures using the
GILDAS\footnote{ http://www.iram.fr/IRAMFR/GILDAS } software. 
The data were calibrated using the IRAM package Continuum and Line Interferometer Calibration (CLIC).  The standard pipeline reduction and
calibration was followed to a large extent, only some poor data scans
were flagged and the use of the standard flux calibrator had to be
enforced in one observing run.  From the resulting uv tables, a
continuum table was produced combining all non-line channels in both
sidebands with the tasks {\it uv$\_$cont} and {\it uv$\_$merge}.  For
the line data, a constant baseline was subtracted from the calibrated
uv tables in the USB. We then reduced the table size by extracting
only the channels with line emission and channels nearby. Finally, we
produced uv tables with four different frequency resolutions 
using the task {\it uv$\_$compress}: the original 2 MHz resolution, 
4 MHz (corresponding to 10.6 \kms), 8
MHz (corresponding to 21.3 \kms), and 12 MHz (corresponding to 31.9 \kms). 

We imaged the data with natural weighting to maximize the sensitivity.
We tested different tapers in order to search for faint, extended
emission, but found no evidence for it.  We will therefore use the
untapered data (for the beam sizes see
Table~\ref{tab:overview_obs_data}).  We tested different cleaning
  procedures: the robust algorithm CLEAN introduced by
  \citet{hogbom74}, the variant developed by \citet{clark80} and the
  method proposed by \citet[][SDI]{steer84} which represents a
  cleaning algorithm which is better adapted to extended structures.
  We finally selected the data cube cleaned with H\"ogbom for the
  compact source Arp~238 (although no major difference was found when
  using the Clark algorithm), and the cube cleaned with SDI which was
  able to best deal with the extended emission in Arp~142 without
  producing artifacts.  For both objects, we used the recommended
loop-gain of 0.2 and truncation threshold of 0.2 of the primary beam
sensitivity. The velocities in this paper are calculated using the 
optical convention and are relative to the Local Standard of 
Rest reference frame.

\section{Results}
For both Arp~142 and Arp~238 we present integrated CO(1-0) maps,
 compared to Spitzer-IRAC \citep{Xu2010} and HST images
\footnote{The HST data are based observations
made with the NASA/ESA Hubble Space Telescope,
and obtained from the Hubble Legacy Archive, which is a collaboration
between the Space Telescope Science Institute (STScI/NANA), the Space Telescope
European Coordinating Facility (ST-ECF/ESA) and the Canadian Astronomy Data Centre (CAD/NRC/CSA).}, 
in Figure~\ref{fig:CO-IRAC}. The total integrated CO fluxes, together 
with other physical parameters, are presented in Table~\ref{tab:results}.
As shown in Figure~\ref{fig:CO-IRAC}, the CO in Arp~142 is widely distributed within a highly  distorted disk of NGC~2936. 
While the CO traces quite well the star-formation as traced by Spitzer
(the red areas in Figure~\ref{fig:CO-IRAC}b, which also coincide with 
the dust lanes in the Hubble images), there is no CO concentration
in the nucleus (Region~2 in Figure~\ref{fig:CO-IRAC}b). 
The total CO flux measured in the NOEMA map of Arp~142 
(Table~\ref{tab:results}) is a
factor of 1.35 higher than that detected by IRAM-30m ($\rm 175.3\pm
2.8\; Jy\; km\; s^{-1}$, \citealp{Lisenfeld2019}), because the NOEMA
measurement covers a significantly larger area than the IRAM beam
(FWHM=$22''$). Assuming the standard conversion factor 
$\rm \alpha_{CO} = 3.2$ $\rm M_\sun K^{-1} km^{-1} s\; pc^2$ \citep{Bolatto2013},
the total molecular gas mass of Arp~142 is 
$\rm M_{mol}=10^{10.29\pm 0.02}\; M_\sun$. This is
about a factor of 2 higher than that of \citet{Bothwell2014} 
based on a CO(2-1) observation of APEX, which has a beam (27\arcsec) 
significantly smaller than the size of the CO emission 
(Figure~\ref{fig:CO-IRAC}a). On the other hand,
the $\rm M_{mol}$ is 0.39 dex lower than that estimated by \citet{Lisenfeld2019},
suggesting that the large aperture correction ($\rm f_{aper}=3.16$)
adopted by \citet{Lisenfeld2019} might have been over-estimated.

\begin{table*}
\caption{NOEMA results and other physical parameters of Arp~142 and Arp~238}
\label{tab:results}
\begin{tabular}{lcccl}
\noalign{\smallskip} \hline \noalign{\medskip}
 & {\bf NGC~2936} & \ \ {\bf Arp~238-E}\ \ & \ \ {\bf Arp~238-W}\ \  & {\bf reference} \\
 & {\bf \ \ (in Arp~142)\ \ } &  &  & \\
 \noalign{\smallskip} \hline \noalign{\medskip}
$I_{\rm CO(1-0)}\; {\rm [Jy\; km\; s^{-1}]\; ^{(a)}}$  &
   $237.4\pm 12.4 $ & $47.0\pm 2.4$ & $27.5\pm 1.4$ & this~work \\
$f_{\rm 2.6mm}\; {\rm [mJy]}$  &
 ......   & $3.9\pm 1.0$ & ......  & this~work \\
optical redshift  &
      0.02331        &   0.03078     & 0.03080    & \citet{Domingue2009} \\ 
luminosity distance [Mpc]  &
      104.0          &   134.1     & 134.1    & \citet{Cao2016} \\ 
kpc per arcsec  &
      0.49          &   0.63    & 0.63    &  \\ 
$\rm log(M_{mol})\; [M_\sun]\; ^{(b)}$  &
  $10.29\pm 0.02$           &   $9.84\pm 0.02\; ^{(c)}$   & $9.70\pm 0.02\; ^{(d)}$ & this work \\ 
$\rm log(M_{HI})\; [M_\sun]$&
 $9.70\pm 0.05$  & \multicolumn{2}{c}{$9.70\pm 0.07\; ^{(e)}$}   & \citet{Zuo2018}, \citet{Huchtmeier1989} \\ 
$\rm log(M_{gas})\; [M_\sun]\; ^{(f)}$&
 $10.39\pm 0.02$  & \multicolumn{2}{c}{$10.23\pm 0.04$}   & this work \\ 
$\rm log(M_{dust})\; [M_\sun]$  &
      $8.37\pm 0.09$    &   $7.81\pm 0.09$   & $7.80\pm 0.09$   & \citet{Cao2016} \\ 
$\rm M_{dust}/M_{gas}$&
 $0.0095\pm 0.0024$  & \multicolumn{2}{c}{$0.0076\pm 0.0017$}   & this work \\ 
$\rm log(M_{star})\; [M_\sun]$  &
      11.16          &   10.80   & 10.62   & \citet{Cao2016} \\ 
$\rm log(L_{IR})\; [L_\sun]$  &
      10.96          &   11.71   & 10.84  & \citet{Cao2016} \\ 
$\rm SFR\; [M_\sun\; yr^{-1}]\; ^{(g)}$  &
      9.83         &  55.30     & 7.59   & \citet{Cao2016} \\ 
\noalign{\smallskip} 
\hline \noalign{\medskip}
\end{tabular}

{\small
\noindent{\bf Notes:} 
$^{(a)}$ Each flux error is estimated as the sum of the rms error 
and a 5\% error including uncertainties of the background subtraction,
aperture setting, and the calibration.
$^{(b)}$ The standard conversion factor 
$\rm \alpha_{CO} = 3.2$ $\rm M_\sun K^{-1} km^{-1} s\; pc^2$ \citep{Bolatto2013} is
adopted.
$^{(c)}$ A missing flux correction factor of 1.10, 
estimated from the comparison with
the IRAM single dish result of \citet{Lisenfeld2019}, is applied
to the NOEMA CO flux. 
$^{(d)}$ The missing flux correction factor is 1.35.
$^{(e)}$ The HI observation includes both Arp~238-E and Arp~238-W. 
$^{(f)}$ The total cold gas mass $\rm M_{gas}=M_{mol}+M_{HI}$.
$^{(g)}$ Assuming $\rm SFR\; (M_\sun\; yr^{-1})
=1.086\times 10^{-10}\times (L_{IR}/L_\sun)$.
}
\end{table*}

\begin{table*}
\caption{Properties of Regions of Interest in Arp~142 and Arp~238 $^{(a)}$}
\label{tab:regions}
\begin{tabular}{lccccccc}
\noalign{\smallskip} \hline \noalign{\medskip}
  {\bf Name} & RA \& Dec & size$^{(b)}$  & $I_{\rm CO(1-0)}$ & $\rm log(M_{mol})^{(c)}$ &
$\rm log(L_{8\mu m,dust})^{(d)}$ & log(SFR)$^{(e)}$ & log(SFE$^{(f)}$) \\
        & (J2000) & [kpc]  & $\rm [Jy\; km\; s^{-1}]$  & $\rm [M_\sun]$  & $\rm [L_\sun]$ & $\rm [M_\sun\; yr^{-1}]$ & $\rm [yr^{-1}]$ \\ 
\\
\noalign{\smallskip} \hline \noalign{\medskip}
 \multicolumn{8}{l}{{\bf Arp~142:}} \\
 Region~1$^{(g)}$ &  09:37:44.97~+02:45:34.4  &  4.2 & 31.71$\pm$ 0.86 &  9.42 & 9.98 &  0.68 & -8.74  \\
 Region~2$^{(h)}$ &  09:37:44.14~+02:45:39.3 &  4.2 & 40.17$\pm$ 0.97 &  9.52 &  9.64 &  0.34 & -9.18 \\
 Region~3$^{(i)}$ &  09:37:43.61~+02:45:43.0 & 4.2 & 49.20$\pm$ 1.07 &  9.61 &  9.43 &  0.12 & -9.49 \\
 Region~4$^{(j)}$ &  09:37:43.92~+02:45:25.8 & 18.0 &246.6$\pm$ 2.40 & 10.31 & 10.11 &  0.81 & -9.50 \\
\\
\multicolumn{8}{l}{{\bf Arp~238:}} \\
  E$^{(k)}$ &13:15:34.93~+62:07:29.1 &  5.4 & 38.43$\pm$ 1.15 &  9.72 & 11.00 &  1.69 & -8.03  \\
  W$^{(l)}$ &13:15:30.70~+62:07:45.2 &  5.4 & 22.68$\pm$ 0.90 &  9.49 & 10.42 &  1.11 & -8.38  \\
\\
\hline \noalign{\medskip}
\end{tabular}

{\small
\noindent{\bf Notes:} 
$^{(a)}$ 
See Panels b) and d) of Figure~\ref{fig:CO-IRAC} for definitions of the regions.
$^{(b)}$ For circular regions: $\rm size\; =\; diameter$;
for irregular regions: $\rm size\; =\; 2\times \sqrt{area/\pi}.$
$^{(c)}$ Assuming 
$\rm \alpha_{CO} = 3.2$ $\rm M_\sun K^{-1} km^{-1} s\; pc^2$ \citep{Bolatto2013}.
$^{(d)}$ Derived from $\rm f_{8\mu m,dust} = f_{8\mu m} - 0.232\times f_{3.6\mu m}$, 
where  $\rm f_{8\mu m}$ and $\rm f_{3.6\mu m}$ are flux densities 
in units of Jy \citep{Helou2004}. These are measured 
after convolving the original
Spitzer-IRAC 8$\mu m$ image (beam-size (FWHM) =2.0\arcsec) 
and 3.6$\mu m$ image (beam-size (FWHM) =1.8\arcsec) to the NOEMA resolutions. 
$^{(e)}$ Assuming $\rm L_{IR} = 4.55\times L_{8\mu m,dust}$ \citep{Shivaei2017}, and
$\rm SFR\; (M_\sun\; yr^{-1}) =1.086\times 10^{-10}\times (L_{IR}/L_\sun)$ \citep{Cao2016}.
$^{(f)}$ $\rm SFE=SFR/M_{H_2}$.
$^{(g)}$ Off-nucleus starburst region \citep{Mora2019}.
$^{(h)}$ Nuclear region of NGC~2936.
$^{(i)}$ Region of the peak of the CO(1-0) emission.
$^{(j)}$ Main disk of NGC~2936. The size of this irregular region is 
       defined to be equal to $\rm 2\times \sqrt{area/\pi}$.
$^{(k)}$ Central region of Arp~238-E.
$^{(l)}$ Central region of Arp~238-W.
}
\end{table*}

\begin{figure*}[!htb]
\centering
\includegraphics[width=0.9\textwidth]{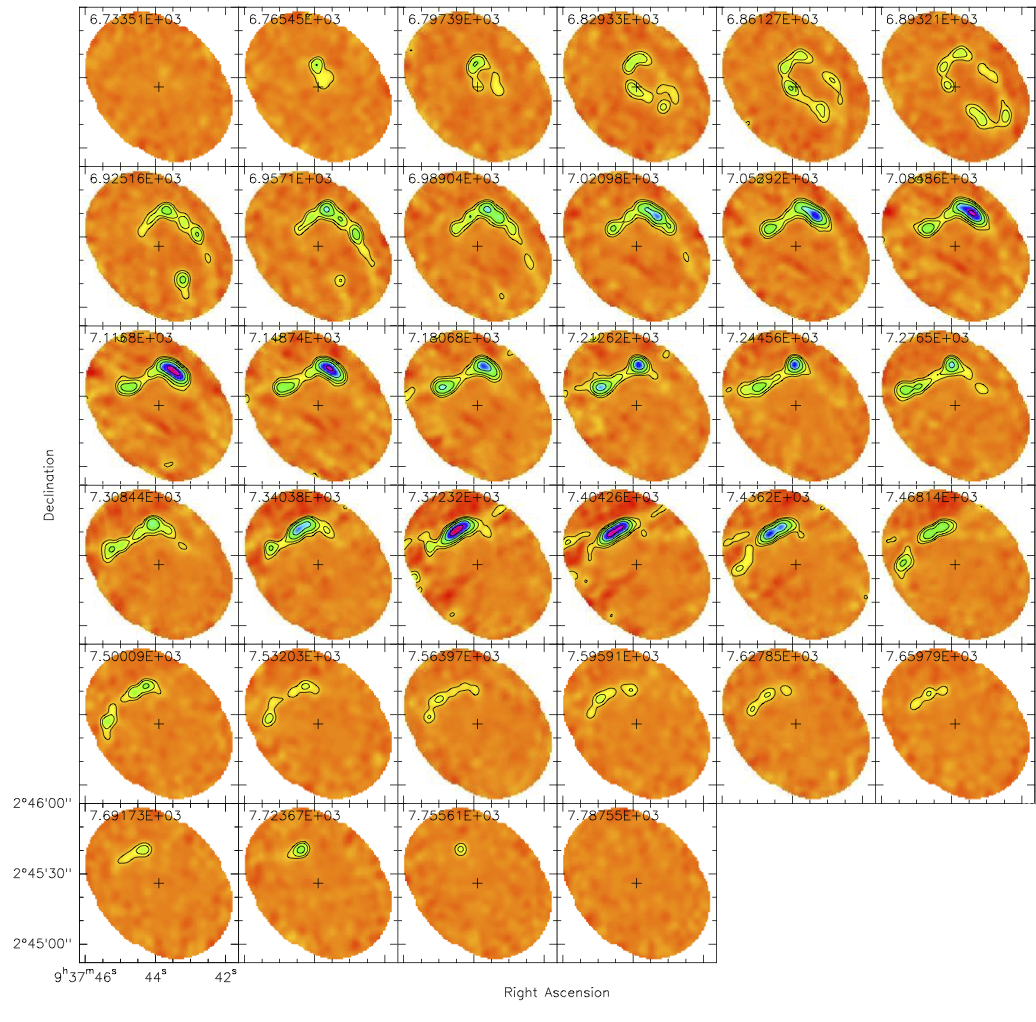}
\caption{CO(1-0) channel maps 
of Arp~142. The bandwidth of the channels is 
$\rm 31.9\; km\; s^{-1}$. The central 
velocity of each channel is marked in the corresponding map. The contour levels are at 6.9, 13.8, 27.6, 55.2, and 110.4 $\rm mJy\; beam^{-1}$.}
\label{fig:Arp142-channels}
\end{figure*}

\begin{figure}[!htb]
\includegraphics[width=0.5\textwidth]{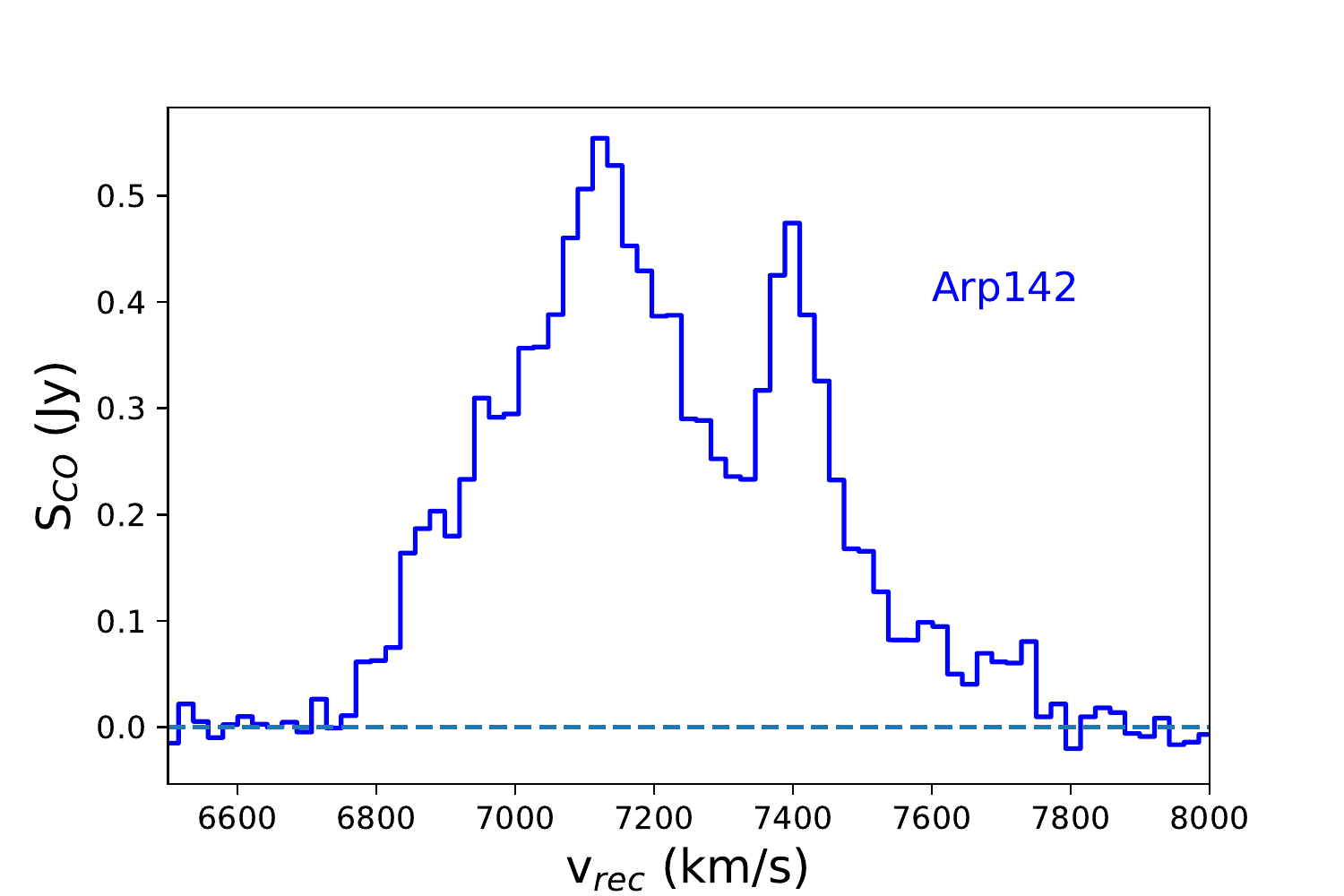}
\caption{CO spectrum of integrated emission of Arp~142 
(within the white ellipse shown in Fig. 1a).}
\label{fig:Arp142-spectrum}
\end{figure}

\begin{figure}[!htb]
\includegraphics[width=0.5\textwidth]{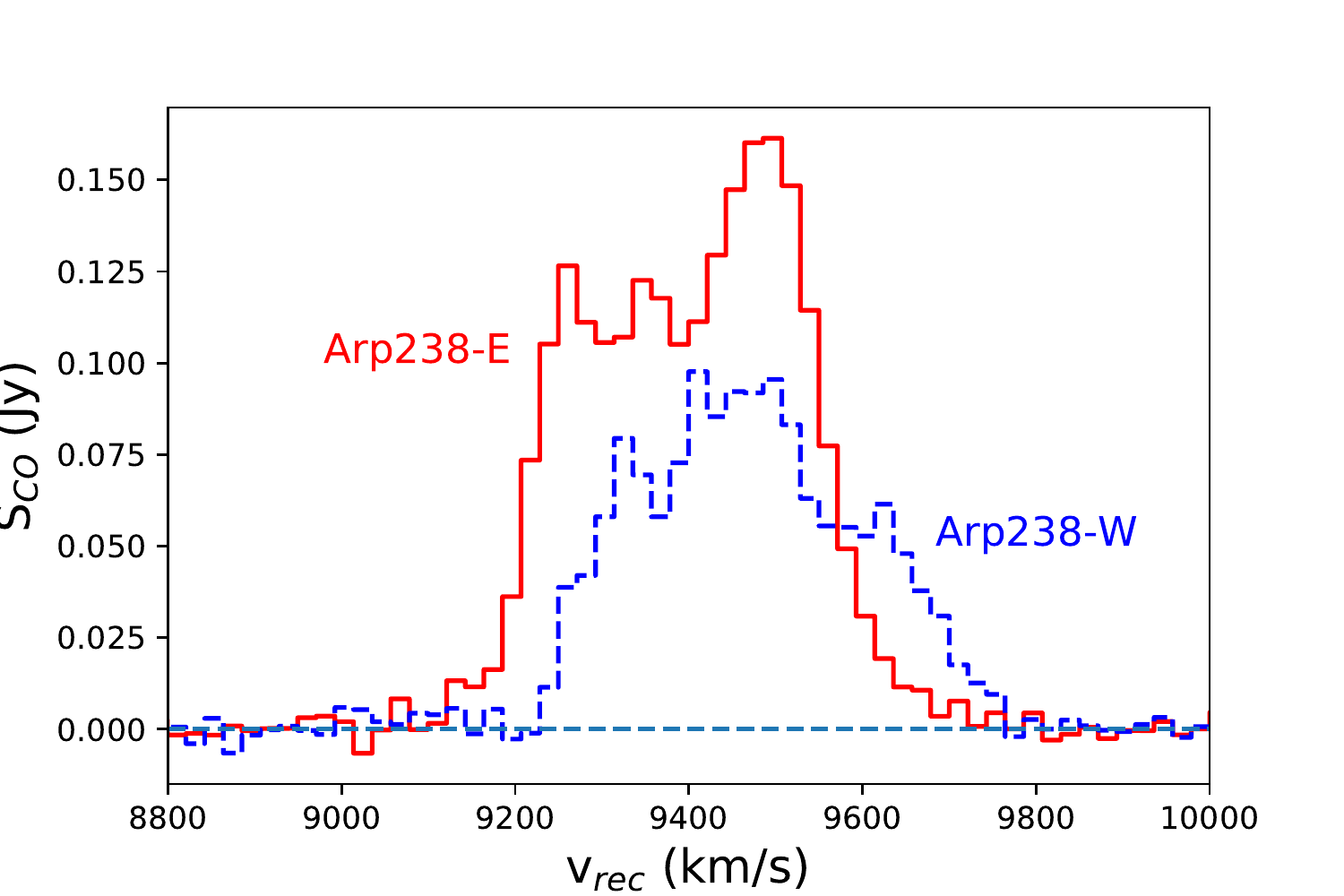}
\caption{CO Spectra of {\bf integrated} emissions of two galaxies in Arp~238
(within the white circles shown in Fig. 1c).}
\label{fig:Arp238-spectra}
\end{figure}

For Arp~238, CO is detected only in two compact sources at the centers
of the two galaxies. Both sources can be fitted well with
  elliptical 2-D Gaussian functions of $\rm FWHM =
  4.4$\arcsec$\times$3.8\arcsec with $\rm P.A. = 80\arcdeg$ (Arp~238-E) and
  and $\rm FWHM = 4.4$\arcsec$\times$3.3\arcsec with $\rm P.A. =
  90\arcdeg$ (Arp~238-W), respectively. Neglecting the small
  difference between the P.A. of each source and
  that of the beam ($\rm P.A. = 88\arcdeg$, Table~\ref{tab:overview_obs_data}), 
  the intrinsic sizes of these sources after
  the deconvolution are 2.8\arcsec$\times$2.3\arcsec (Arp~238-E) and
  2.8\arcsec$\times$1.3\arcsec (Arp~238-W), respectively.  These are
  smaller than the NOEMA beam size (see
  Table~\ref{tab:overview_obs_data}), and slightly larger than the
  sizes of radio continuum sources ($\sim 1$\arcsec -- 2\arcsec) detected in
  the high resolution VLA maps of \citet{Condon1991}.  
No diffuse CO is detected in Arp~238. Compared to their CO fluxes measured by
IRAM-30m ($\rm 51.8\pm 3.4\; Jy\; km\; s^{-1}$ and $\rm 37.0\pm 3.3\;
Jy\; km\; s^{-1}$), the NOEMA observations missed 9\% and 26\% of the
CO emissions of Arp~238-E and Arp~238-W, respectively. The missed
emissions are most likely associated with diffuse gas in the disks
and/or tidal tails.  The continuum was detected only at the
  nucleus of Arp~238-E as a point source.  No continuum emission was
detected in Arp~142, nor in Arp~238-W.

Figure~\ref{fig:Arp142-channels} shows the channel maps of
Arp~142. Interestingly, there appears to be an off-centric ring-like
structure in the channel maps of $\rm v < 7000\; km\; s^{-1}$,
most clearly visible in the channel of $\rm v = 6893\;
km\; s^{-1}$. It seems to grow in size from $\rm v =6765
\; km\; s^{-1}$ to $\rm v =6957\; km\; s^{-1}$. This could
be an expanding cone-like structure in 3-D with the top of the cone
moving toward the observer (i.e. blue-shifted), which is 
also the direction of the motion of the elliptical companion ($\rm z=0.02265$,
corresponding to $\rm v=6795\; km\; s^{-1}$). Notably, most
CO of $\rm v > 7000\; km\; s^{-1}$ is in the north, and the
channel maps are consistent with a highly distorted rotating disk.

The spectrum of the integrated emission of Arp~142 is presented in 
Figure~\ref{fig:Arp142-spectrum}. It shows an asymmetric double-horn shape,
representing a distorted disk. As shown in channel maps 
(Figure~\ref{fig:Arp142-channels}), the ring-like structure resides in
the blue peak which has a FWHM of $\rm \sim 300\; km\; s^{-1}$, 
significantly broader than the red peak ($\rm FWHM \sim 100\; km\; s^{-1}$).
The total spectrum is very wide with a 
full width at zero intensity (FWZI) of $\rm \sim 1000\; km\; s^{-1}$, similar to 
that of Arp~118 which is a LIRG and a ring galaxy with extremely high CO
luminisity \citep{Gao1997}.

Because neither of the two galaxies in Arp~238 is well resolved, we show only
the spectra of their integrated emissions in
Figure~\ref{fig:Arp238-spectra}. Both sources have broad (FWHM$\rm \sim
400\; km\; s^{-1}$) and multi-peak line profiles, suggesting complex
kinematics within them. This is consistent with the strong interaction
between them revealed by the long tidal tails shown in optical maps,
and with their high SFR which can cause strong turbulence and feedbacks.

In order to investigate how the SFE varies within the two pairs,
  we inspect several regions in Arp~142 (defined in
  Figure~\ref{fig:CO-IRAC}b) and the two central regions in Arp~238
  (Figure~\ref{fig:CO-IRAC}d).  The results are listed in
  Table~\ref{tab:regions}.  For Arp~142, the main disk (Region~4) has
  a rather low SFE, 0.43 dex lower than the mean of AMIGA control
  sample of \citet{Lisenfeld2019} which is $\rm log(SFE/yr^{-1}) =
  -9.07\pm 0.05$. Interestingly, there is a significant variation of
  SFE within Arp~142.  The CO-peak region (Region~3) has a similarly
  low log(SFE) as that of Region~4, and log(SFE) of the nuclear region
  (Regino~2) is comparable to the mean of the control sample.  On the
  other hand, the starburst region (Region~1) near the eastern tip of
  the distorted disk has an SFE more than 0.7 dex higher than that of
  Region~3, showing a moderate SFE enhancement ($\sim 0.3$ dex)
  compared to the control sample. However, only $\sim 10\%$ of
  molecular gas in Arp~142 is found in Region~1, the majority of the
  remaining gas is likely to have relatively low SFE, as suggested by
  the result of Region~4.  In contrast, the two central regions in
  Arp~238 dominate the total $\rm M_{mol}$ in Arp~238, and both have
  very high SFEs. Compared to the control sample, Arp~238-E shows an
  SFE enhancement of more than 1 dex whereas Arp~238-W has an
  enhancement of $\sim 0.7$ dex.

\begin{figure}[!htb]
\includegraphics[width=0.5\textwidth]{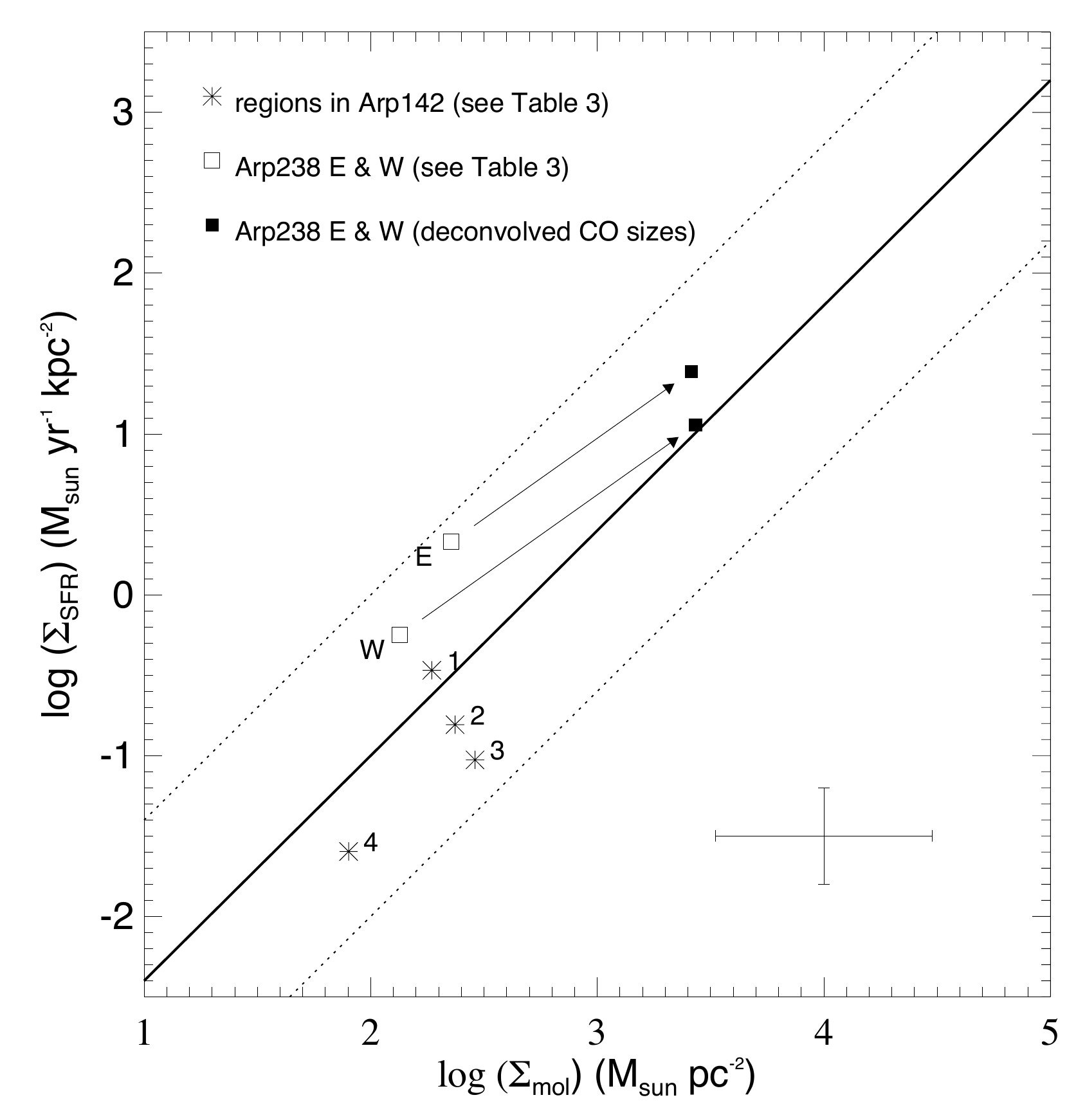}
\caption{Kennicutt-Schmidt (K-S) plot of regions in Arp~142 and
Arp~238 (see Table~\ref{tab:regions} for their definitions). For the two 
central regions in Arp~238 (E and W), the black filled squares show the
estimated $\rm \Sigma_{SFR}$ and $\rm \Sigma_{mol}$ values 
when the sizes in Table~\ref{tab:regions}
are replaced by the deconvolved sizes of the corresponding 
CO sources. 
Black solid line shows the K-S relation \citep{Kennicutt1998b} 
converted to the Kroupa IMF, and the two dotted lines mark the limits of
10 times deviation. 
The large black plus sign in the bottom right corner illustrates error bars of
$\rm \Sigma_{SFR}$ (a factor of 2) and of $\rm \Sigma_{mol}$ (a factor of 3),
dominated by systematic uncertainties in the $\rm L_{8um,dust}$-to-SFR and
$\rm L_{CO}$-to-$\rm M_{mol}$ conversions, respectively.
}
\label{fig:KS-plot}
\end{figure}

In Figure~\ref{fig:KS-plot} the SFR surface densities ($\rm
\Sigma_{SFR}$) of regions in Table~\ref{tab:regions} are plotted
against the surface densities of molecular gas ($\rm
\Sigma_{mol}$). They scatter around the standard Kennicutt-Schmidt relation
(hereafter K-S relation; \citealt{Kennicutt1998b}), with most of regions in
Arp~142 found below the K-S relation and the two central regions in
Arp~238 high above it. It is worth pointing out that the CO emission in
Arp~238-E and Arp~238-W is not resolved, and the estimated intrinsic
sizes of the two sources are much smaller than that of the two central
regions. Also, observations of various SFR indicators in the radio
continuum \citep{Condon1991}, in 8$\mu m$ Spitzer-IRAC band \citep{Xu2010},
and in H$\alpha$ line emission \citep{Hattori2004} all show the
dominance of two nuclear sources over the entire pair.  The high
resolution 8.44 GHz VLA maps (beam = 0.25\arcsec) of
\citet{Condon1991} show that the emission regions associated with
Arp~238-E and Arp~238-W have sizes of $\sim 2$\arcsec and $\sim
1$\arcsec, respectively. These are even smaller than the estimated
intrinsic sizes of the CO emission regions
(2.8\arcsec$\times$2.3\arcsec for Arp~238-E,
2.8\arcsec$\times$1.3\arcsec for Arp~238-W). Assuming that both CO and
8$\mu m$ emissions in Arp~238-E are from an elliptical region of size
2.8\arcsec$\times$2.3\arcsec and those in Arp~238-W from a region of
size 2.8\arcsec$\times$1.3\arcsec, we raise both $\rm \Sigma_{SFR}$
and $\rm \Sigma_{mol}$ of Arp~238-E by a factor of 11.5 and $\rm
\Sigma_{SFR}$ and $\rm \Sigma_{mol}$ of Arp~238-W by a factor of
20.3. This moves the data points representing the two central regions
to the upper-right part of the K-S plot and,
because of the non-linearity of the K-S relation ($\rm power=1.4$), 
much closer to the line of the K-S relation.

There is a large uncertainty for the conversion factor 
$\rm \alpha_{CO}$, in particular for (U)LIRGs \citep{Bolatto2013,Downes1998}.
We compare the molecular gas mass estimated using the adopted
$\rm \alpha_{CO}$ with the HI mass \citep{Zuo2018} 
and dust mass \citep{Cao2016} for Arp~142 and Arp~238,
in order to constrain the effect of this uncertainty to our results.
As shown Table~\ref{tab:results}, 
the dust-to-gas ratios of the two pairs are $0.0095\pm
0.0024$ and $0.0076\pm 0.0017$, both consistent with the value
($\sim 0.007$) of \citet{Draine2007} obtained for local
spiral galaxies. This suggests that globally our results are not significantly
affected by the uncertainty of $\rm \alpha_{CO}$. 
It is worth noting that \citet{Lisenfeld2019} made
an in-depth discussion on the applicability of the standard conversion
factor to a sample of close major-merger pairs, to which both Arp~142
and Arp~238 belong, and concluded that their results on molecular mass
derived using $\rm \alpha_{CO} = 3.2$ $\rm M_\sun
K^{-1} km^{-1} s\; pc^2$ are robust. Nevertheless, conservatively, we
put an error bar of a factor of 3 for $\rm \Sigma_{mol}$
\citep{Renaud2019} in Figure~\ref{fig:KS-plot}, applicable to all data
points in the plot. Also plotted is an error bar of a factor of 2 for
$\rm \Sigma_{SFR}$ (dominated by systematic uncertainties in the $\rm
L_{8\mu m,dust}$-to-SFR conversion).

\section{Discussion}

Given the peculiar optical morphology of Arp~142 and Arp~238, it
  is likely that both pairs have undergone strong interactions
  recently, and their very different molecular gas distributions and
  SFEs found in NOEMA observations are due to differences in their
  interactions.  The dynamic histories of both Arp~142 and Arp~238
were simulated by \citet{Holincheck2016} using a simple three-body
simulation code, and their best models found that both pairs have
undergone close encounters recently: For Arp~142, the pericentric
  passage has $\rm r_{min}= 8.95\pm 1.14$ kpc and occurred 
  $\rm t_{min}= 78.0\pm 6.3$ Myr ago, whereas for Arp~238 $\rm r_{min}=
  12.54\pm 2.95$ kpc and $\rm t_{min}=58.9\pm 13.1$ Myr. On the other hand,   
the two interacting systems
have diagonally different orbital orientations: while the interaction is coplanar for both galaxies in Arp~238, it is nearly
perpendicular for the spiral galaxy in Arp~142.  
Also, the radial
  velocity difference between the two galaxies in Arp~142 is
  201~km~$\rm s^{-1}$, significantly higher than that of Arp~238 which is
  only 6~km~$\rm s^{-1}$.  \citet{Mora2019} found a best-match model for
  Arp~142 in the suite of more sophisticated Galmer SPH simulations
  (\citealp{DiMatteo2008}; \citealp{Chilingarian2010}),
which agrees very well with that of \citet{Holincheck2016}, with $\rm r_{min}= 8\;
kpc$, $\rm v_{min}=300\; km\; s^{-1}$, $\rm t_{min}= 52\pm 25$ Myr,
and the disk of the spiral galaxy perpendicular to the orbit plane.

Thus, simulation results indicate that Arp~142 and Arp~238 have
  gone through very different types of interactions: the former a high-speed 
  ($\rm \sim 300\; km\; s^{-1}$) head-on collision between the disk
  and the elliptical companion, and the latter a low-speed coplanar interaction
  between two spiral galaxies. Both observations and simulations 
(\citealt{Theys1977}; \citealt{Appleton1987, Appleton1996})
have shown that off-centric high-speed head-on
collision produces ring-like density waves expanding through
both stellar and gaseous disks, and pushing 
gas in the central region to the
outer disk. This scenario is consistent with the NOEMA data of Arp~142.
Meanwhile, shocks and turbulence associated with the ring can
either compress gas clouds and trigger starbursts similar to that in
Arp~142 (\citealp{Gao1997}; \citealp{Lamb1998}; \citealp{Higdon2011};
\citealp{Renaud2018}), or inject kinematic energy into clouds and 
stabilize them
against collapse \citep{Alatalo2014, Guillard2012}.  Whether this
can explain the large variation of the SFE in Arp~142 will be
the subject of a follow-up study of the kinematics and its
relation to SFE in Arp~142, via a high resolution hydrodynamic
simulation (Renaud et al. in preparation).

On the other hand, the very high SFEs of Arp238-E and Arp~238-W are
apparently related to the compactness of the starbursts in their
nuclei, which have $\rm \Sigma_{mol}$ approaching those of ULIRGs
\citep{Scoville1991}.  This can be explained by simulations of
\citet{Barnes1996} and \citet{Hopkins2009a}, which predicted that
gravitational tidal torques in low-speed coplanar interactions can
trigger strong gas inflows that lead to nuclear gas concentrations and
nuclear starbursts.

Does the contrast between Arp~142 and Arp~238 represents a common
difference between S+E and S+S pairs?  Namely, do more S+E pairs have
high-speed and high-inclination interactions while low-speed coplanar
interactions are more common in S+S pairs?  A definite answer to this
question can only be obtained through dynamic simulations of a
complete pair sample, which is beyond the scope this paper.  However,
some hints can be found in the following statistics of the H-KPAIR
sample \citep{Cao2016}: For S+E pairs in H-KPAIR,
  the average radial velocity difference between pair members is $\rm
  215.7\pm 20.4\; km\; s^{-1}$, higher than that for S+S which is $\rm
  165.9\pm 18.2\; km\; s^{-1}$. Also, for  S+E pairs and S+S pairs, the means of
  the number of galaxies of $\rm M_r \leq -19.5$ found within 1~Mpc projected
  radius from the pair center and with redshift differences (compared
  to that of the pair) $\rm < 500\; km\; s^{-1}$ are $5.14\pm 0.59$, and
  $3.81\pm 0.35$, respectively. This indicates 
  that, compared to S+S pairs, S+E pairs
  are in higher local density environment and therefore more likely
  found in groups or clusters. 
While isolated pairs formed in
IGM filaments may preferentially have coplanar orbits, pairs in
groups/clusters are likely to have significantly disturbed orbits that
are more randomly oriented, as suggested by the results of
\citet{Dubois2014}.  Indeed, Arp~142 itself is in a group with 6
members brighter than $\rm M_r = -19.5$, and NGC~2936 is the
first-ranked member of the group \citep{Yang2007}.  It is worth noting
that \citet{Elagali2018} found, in an investigation of results of
EAGLE simulations, that ring galaxies triggered by recent high-inclination
collisions are more likely found in massive groups, and they tend to
have low SFEs.  
These results favor the hypothesis that S+E pairs
  are more likely to have high-speed and high-inclination interactions
  and S+S pairs low-speed coplanar interactions, which may
  result in a lower chance for S+E pairs to have high SFE nuclear
  starbursts compared to S+S pairs.

\section{Summary and Conclusions}

Previous observations of the SFR \citep{Cao2016}
 and the molecular \citep{Lisenfeld2019} 
and atomic \citep{Zuo2018} gas content
of the H-KPAIR sample have shown pronounced differences between 
S+E and S+S pairs: The sSFR is only enhanced 
  in S+S pairs, and there is a significant difference
 between the SFEs of S+E and S+S pairs
\citep{Lisenfeld2019}.
In order to probe the physical mechanism for
these differences, we carried out NOEMA imaging observations of CO(1-0)
line emission in two representative pairs: the S+E pair Arp~142 and
the S+S pair Arp~238.  In the sample of \citet{Lisenfeld2019}, the
spiral component of Arp~142 has the highest SFR among galaxies in S+E
pairs and Arp~238-E (a LIRG) the second highest SFR among galaxies
in S+S pairs, whereas the SFE of the former is about 30 times lower than
that of the latter.  

The NOEMA observations, with a linear resolution of about 2 kpc, gave the following results:

\begin{itemize}
\item The CO emission in Arp~142 is widely
distributed over a highly distorted disk of the spiral galaxy
(NGC~2936) without any nuclear concentration, and an off-centric
ring-like structure is discovered in channel maps.  

\item
There is a significant variation of the SFE within Arp~142. 
The starburst region (Region~1)
near the eastern tip of the distorted disk has an SFE more than 0.7 dex
higher than that of the CO-peak region (Region~3) and shows 
a moderate SFE enhancement 
($\sim 0.3$ dex) compared to the mean of the AMIGA control sample 
of isolated galaxies \citep{Lisenfeld2019}. 

\item Only $\sim
10\%$  of the molecular gas in Arp~142 is found in the starburst region, whereas 
the majority of the remaining gas has  relatively low SFE
as suggested by the result for the main disk (Region~4) which has an SFE
0.43 dex lower than the mean of control sample.

\item In Arp~238, CO is
detected only in two compact sources at the two galactic centers.

\item The two central
regions in Arp~238 dominate the total $\rm M_{mol}$ in Arp~238,
and both have very high SFEs. Compared to the control sample,
 Arp~238-E shows an SFE enhancement of more than 1 dex whereas 
Arp~238-W has an enhancement of $\sim 0.7$ dex.

\end{itemize}
The differences between these two merger pairs are most likely due to 
different orbital parameters of  the encounters:
Simulations in the literature
\citep{Holincheck2016, Mora2019} have found that  Arp~142 has
undergone a high-speed off-centric head-on collision 
while the Arp~238 has gone through a low-speed coplanar interaction.  
The extended CO distribution and large SFE variation in Arp~142
are most likely related to the shocks and turbulence associated with an 
expanding large-scale ring triggered by the head-on collision. 
On the other hand, the very high SFEs of Arp238-E and Arp~238-W are
related to the compactness of the starbursts in their
nuclei which have very high $\rm \Sigma_{mol}$. As predicted by simulations 
\citep{Barnes1996, Hopkins2009a},
gravitational tidal torques in low-speed coplanar interactions can
trigger strong gas inflows that lead to nuclear gas concentrations and
nuclear starbursts.

These differences in orbits might be typical for S+S and S+E pairs in general. 
Statistics for the H-KPAIR sample indicate that on average
S+E pairs have a higher radial velocity difference
and are more likely found in groups or clusters compared to S+S pairs.
Since isolated pairs formed in IGM filaments may
preferentially have coplanar orbits \citep{Dubois2014} and pairs in
groups/clusters are expected to have significantly disturbed orbits
that are more randomly oriented, we propose the following hypothesis
in analog to the NOEMA results for Arp~142 and Arp~238: S+E pairs are more
likely to have high-speed and high-inclination interactions
  and S+S pairs low-speed coplanar interactions, which may
  result in a lower chance for S+E pairs to have high SFE nuclear
  starbursts compared to S+S pairs.

\vspace*{1cm}
\noindent{\it Acknowledgments}:

This work is supported by the National Key R\&D
Program of China No. 2017YFA0402704 and by National Natural Science Foundation
of China (NSFC) No. 11873055, and is sponsored 
(in part) by the Chinese Academy of Sciences (CAS)
through a grant to the CAS South America Center for Astronomy (CASSACA).
CKX acknowledges NSFC grants No. 11733006.
UL acknowledges support by the research
project  AYA2017-84897-P from the Spanish Ministerio
de Economia y Competitividad, from the European Regional Development
Funds (FEDER) and the Junta de Andalucia (Spain) grants FQM108.
YG acknowledges NSFC grants No. 11861131007, 12033004, and 11420101002, and Chinese Academy of Sciences Key Research Program of Frontier 
Sciences (Grant No. QYZDJ-SSW-SLH008).  
FR acknowledges support from the Knut and Alice Wallenberg Foundation.
This work is based on observations carried out under project number 
W19BL with the IRAM NOEMA Interferometer. 
IRAM is supported by INSU/CNRS (France), MPG (Germany) and IGN (Spain).
We thank the IRAM support astronomer, Ka Tat Wong, for valuable help and advice with the reduction of the data.

\bibliographystyle{apj}
\bibliography{/Volumes/SSD/data1/bibliography/papers_biblio}
\end{document}